\documentclass[useAMS,doublespacing]{gGAF2e}

\usepackage{natbib}
\usepackage[latin1]{inputenc}
\usepackage{psfig,epsfig}
\usepackage{amssymb}

\newcommand{\dd}[2]{\frac{{\rm d} #1}{{\rm d} #2}}

\newcommand{\dpart}[2]{\frac{\partial #1}{\partial #2}}
\newcommand{\dpartt}[2]{{\partial #1}/{\partial #2}}

\begin{document}

\title{Time-dependent analytic solutions of quasi-steady shocks with cooling}

\markboth{P. Lesaffre}{Time-dependent analytic solutions of
quasi-steady shocks with cooling} 

\author{PIERRE LESAFFRE$^*\dagger$\thanks{$^*$Email:
pierre.lesaffre@lra.ens.fr} \\ $\dagger$ Laboratoire de
Radioastronomie, 24 rue Lhomond, 75231 PARIS Cedex 05, France}

\received{Received 24 March 2006}

\maketitle

\begin{abstract}
  I present time-dependent analytical solutions of
quasi-steady shocks with cooling, where quasi-steady shocks are
objects composed of truncated steady-state models of shocks at any
intermediate time. I compare these solutions to simulations with a
hydrodynamical code and finally discuss quasi-steady shocks as
approximations to time-dependent shocks. Large departure of both the
adiabatic and steady-state approximations from the quasi-steady
solution emphasise the importance of the cooling history in
determining the trajectory of a shock.\\

\end{abstract}

\section{Introduction}
Analytic solutions of well defined problems are often used as
benchmark tests for hydrodynamical codes. However, the most widely used
tests such as the Sedov blast wave \citep{sedov} or the Sod shock tube
test \citep{sod} are almost all adiabatic or with Mach numbers of
order unity.  By contrast, gas in astrophysical contexts can be
subject to strong cooling and Mach numbers of order a few hundreds are
common place in protostellar jets. There is hence a lack
of analytical solutions with cooling or at high Mach numbers. I present
here time-dependent analytical solutions of quasi-steady shocks with
cooling for arbitrarily high Mach numbers.

\cite{PI} and \cite{PII} studied the temporal evolution of molecular
shocks and found that they are most of the time in a quasi-steady
state, ie: an intermediate time snapshot is composed of truncated
steady state models.  They showed that if the solution to the steady
state problem is known for a range of parameters, it is possible to
compute the time-dependent evolution of quasi-steady shocks by just
solving an implicit ordinary differential equation (ODE). Since the
steady state problem is itself an ODE, it is very easy to {\it
numerically} compute the temporal evolution of quasi-steady shocks.

However, a given cooling function of the gas does not necessarily lead
to an {\it analytical} steady state solution.  Furthermore, even when
it does, the implicit ODE which drives the shock trajectory does not
necessarily have an analytical solution itself. In this paper,
we tackle the problem in its middle part : we assume a functional
form for the steady state solutions (section 2). We then show 
how to recover the underlying cooling function that yields
these steady states (section 3). Finally, we exhibit a case
where the shock trajectory has an analytical solution (section 4)
and we compare it to numerical simulations (section 5). Results are
discussed in section 6 and conclusions drawn in section 7.

\section{Method}

 Consider the following experimental set up: we throw gas with a
 given pressure, density and supersonic speed $v_0$ on a wall.  We
 assume a perfect equation of state with adiabatic index $\gamma$.  We
 assume the net rate of cooling of the gas is a function
 $\Lambda(\rho,p)$ where $p$ and $\rho$ are the local density and
 pressure of the gas. The gas is shocked upon hitting the wall, heated
 up by viscous friction and an adiabatic shock front develops that
 soon detaches from the wall. Behind this front, the gas progressively
 cools down towards a new thermal equilibrium state and a relaxation layer
unrolls.

 All physical quantities are normalised using the entrance values of
 pressure and density, so that the sound speed of the unshocked gas is
 $c_0=\sqrt{\gamma}$. The time and length scales will be later specified
 by the cooling time scale (see section~\ref{cool}).

 Consider now the set of all possible stationary states for the velocity
profile {\it in the frame of the shock front}. A given entrance
speed $u_0$ in the shock front provides the velocity $u$ at a given distance
$y$ behind the shock front:
\begin{equation}
\label{steadyf}
u=f(y,u_0) \mbox{.}
\end{equation}
The adiabatic jump conditions for an infinitely thin (or steady)
shock enforce  $f(0,u_0)=u_a(u_0)$ where
\begin{equation}
u_a(u_0)=\frac{\gamma-1}{\gamma+1}u_0+\frac{2\gamma}{\gamma+1}\frac{1}{u_0}
\mbox{.}  
\end{equation}
Unfortunately, that $f$ is a simple algebraic function of $y$ and $u_0$
 does not necessarily imply an algebraic form for $\Lambda(\rho,p)$. 
It is in fact more appropriate to express $y$ in terms of $u$
and $u_0$. We hence rather write (\ref{steadyf}) in
the following manner:
\begin{equation}
\label{steadyg}
y=g(u,u_0) 
\end{equation}
with the condition 
\begin{equation}
\label{condition}
g(u_a,u_0)=0 \mbox{.}
\end{equation}
Section \ref{cool} details how to recover $\Lambda(\rho,p)$ from $g(u,u_0)$.
  
  \cite{PII} provide the ODE for the evolution of the distance from
  the shock to the wall $r$ with respect to time $t$ if the shock is
  quasi-steady at all times:
\begin{equation}
\label{trajf}
\dot{r}=f(r,\dot{r}+v_0)
\end{equation}
where a dot denotes a derivative with respect to time. This equation
can also be expressed with the function $g$:
\begin{equation}
\label{trajg}
r=g(\dot{r},\dot{r}+v_0) \mbox{.}
\end{equation}
In section \ref{shock} I show how one can integrate equation (\ref{trajg})
and give an analytical expression for a simple form of $g$.
In section \ref{num} I compare this solution to a time-dependent
numerical simulation.

\section{Cooling function}

\subsection{General procedure}
\label{cool}
 Let us write the equation of steady-state hydrodynamics in the frame 
of the shock with entrance parameters $(\rho,p,u)=(1,1,u_0)$:
\begin{equation}
\label{smass}
\rho u=u_0 \mbox{,}
\end{equation}
\begin{equation}
\label{smom}
p+\rho u^2=1+u_0^2
\end{equation}
and
\begin{equation}
\label{lurp}
\dpart{}{y}[u(\frac{\gamma}{\gamma-1}p+\frac12 \rho u^2)]=\Lambda 
\end{equation}
with the boundary condition $u=u_a(u_0)$ at $y=0$. 
  
One can solve the equations (\ref{smass}) and (\ref{smom}) for $\rho$ and $p$
and use the relations into (\ref{lurp}) which becomes:
\begin{equation}
\Lambda(u,y,u_0)=\dpart{}{y}[u(\frac{\gamma}{\gamma-1}(1-u u_0+ u_0^2)+\frac12
u u_0)]
\end{equation}

Expansion of the derivative with respect to $y$ provides:
\begin{equation}
\label{Luup}
\Lambda(u,u',u_0)=\frac{\gamma}{\gamma-1}(1+u_0^2)u'
-\frac{\gamma+1}{\gamma-1}u_0 u u'
\end{equation}
where $u'=\dpartt{u}{y}$.

By taking the derivative of equation (\ref{steadyg}) we easily extract
$u'$ in terms of $u$ and $u_0$:
\begin{equation}
\label{uprime}
u'=1/\dpart{g}{u}(u,u_0)\mbox{.}
\end{equation}
(\ref{uprime}) combined with (\ref{Luup}) provides $\Lambda(u,u_0)$.
(\ref{smass}) and (\ref{smom}) finally give $\Lambda(\rho,p)$.

\subsection{First application}
  I now illustrate this method with a simple function $g(u,u_0)$. In a
  typical radiative shock in the interstellar medium, the post-shock
  gas finally cools down to nearly the same temperature as the
  pre-shock gas. To mimic this effect, we need a cooling function such
  that its thermal equilibrium is the isothermal state. In other
  words, $\Lambda(\rho,p)=0$ implies $\rho=p$. This is equivalent to
  asking that the final steady velocity of any shock verifies the
  isothermal jump condition $u=u_i$ where $u_i(u_0)=1/u_0$:
\begin{equation}
\label{cond2}
\lim_{u \rightarrow u_i}\dpart{g}{u}(u,u_0)=-\infty \mbox{.}
\end{equation} 
 To verify both conditions (\ref{condition}) and (\ref{cond2}) we
take the simple form:
\begin{equation}
\label{gfunc}
g(u,u_0)=\beta\frac{u_a(u_0)-u}{u-u_i(u_0)}
\end{equation}
where $\beta>0$ determines the strength of the cooling. Setting a length
scale allows to assume $\beta=1$.
The above procedure (section \ref{cool}) yields:
\begin{equation}
\label{lnum}
\Lambda(p,\rho)= -\frac{1}{\beta}\left( \frac{\left( 1 + \gamma \right) \,
       \left[ 1 + p\,\left( \gamma\,\rho-\gamma-1 \right)  \right] \,
       {\left( p - \rho \right) }^2}
{
       {\left(  \gamma-1 \right) }^2\,
       {\left(  \rho-1  \right) }^{\frac{3}{2}}\,
       \left(  p\,\rho -1\right) \,
       \sqrt{\rho\,\left(  p-1 \right) }} \right) 
\end{equation}

\begin{figure}
\centerline{ \psfig{file=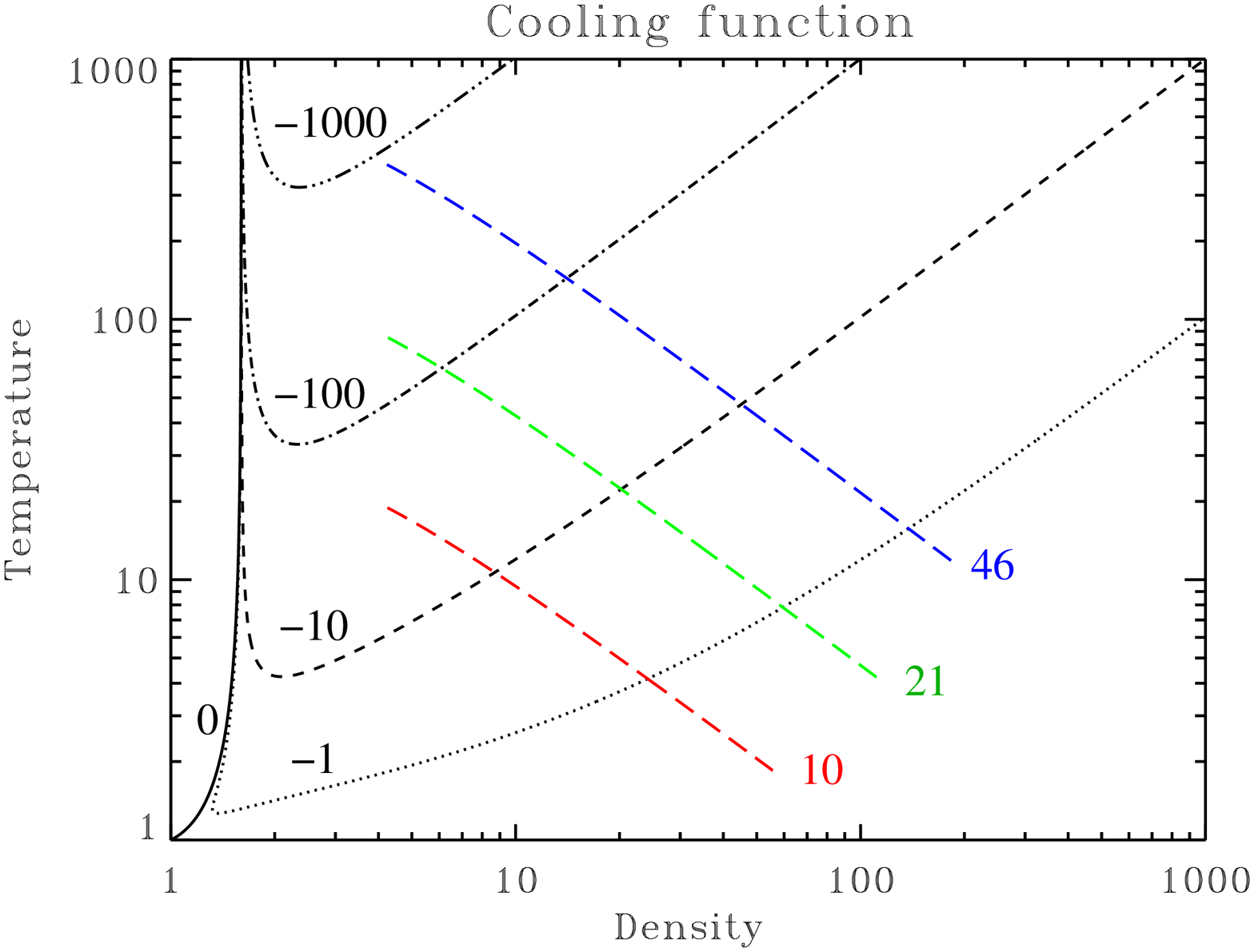,width=9cm,angle=-90} }
\caption{Contour plot of the cooling function $\Lambda$ defined by
equation (\ref{lnum}) for $\beta=1$ and $\gamma=5/3$ with respect to
temperature $T=p/\rho$ and density $\rho$. Levels for the contours are
0 (solid line) and respectively -1, -10, -10$^2$ and, -10$^3$ for
dotted, dashed, dash-dotted and dash-dot-dot-dotted lines. Overlayed
(long-dash) are three typical shock thermal profiles at age $t=100$
(from bottom to top $v_0=10$, 21 and 46).}
\label{lambda2}
\end{figure}

This cooling function is displayed on figure \ref{lambda2}. In
addition to the temperature $T=p/\rho=1$ solution, the thermal
equilibrium state $\Lambda(p,\rho)=0$ is also realised when the factor
$\left[ 1 + p\,\left( \gamma\,\rho-\gamma-1 \right) \right]$ is set to
zero.  However, this state is practically never achieved in the
relaxation layer of a shock as it happens for densities
$\rho<1+1/\gamma$ and $\rho$ is always greater than the adiabatic
compression factor $C_a=(\gamma+1)/(\gamma-1+2\gamma u_0^{-2})\simeq
1+2/(\gamma-1)$ for strong shocks. In the high temperature limit
$\Lambda$ scales like $T^{\frac32}$ which is reminiscent of the
collisional coupling between gas and dust for low dust
temperatures. But in the high density limit, $\Lambda\simeq
\rho^{-\frac12} T^{\frac32}$ which yields a rather unphysical scaling
on the density.  In the next subsection, I show how to improve the
physical relevance of the cooling function, at the loss of the
analytical integrability of the trajectory.

\subsection{Second application}
\label{apcool}

In this section, I briefly illustrate how one can obtain semi-analytic
approximation of shocks for any kind of cooling function. I start with
a given cooling function $\Lambda_0$ and compute an analytical approximation
to the steady state function $g(u,u_0)$. I then recover the underlying
cooling function $\Lambda_1$ for this approximate steady state and
check how $\Lambda_1$ and $\Lambda_0$ differ.

A very common form for the cooling due to a collisionally excited
(unsaturated) line is 
\begin{equation}
\Lambda_0=-\frac1{\beta}\rho^2 \exp[{-T_0/T}]
\end{equation}
where $T_0$ is the temperature of the transition and $\beta$ scales the
strength of the cooling. I use $\beta=1$ for simplification (it amounts
to specify a length scale without loss of generality).

If we apply the procedure of section \ref{cool} {\it backwards} we
have to integrate
\begin{equation}
\label{statcool}
\dd{y}{u}=\frac{u^2 \left[ \gamma (1+u_0^2) - (1+\gamma) u u_0 \right]}
{(\gamma-1)\,u_0^2}\, \exp[{\frac{T_0\,u_0}{u\,(1+u_0^2-u\,u_0)}}]
\end{equation}
to find the equation for the stationary velocity profile. This
equation does not have an analytical solution. But there are many
approximations to the right hand side that will allow to treat the
problem. 

For example, we can simplify (\ref{statcool}) by using the strong
shock approximation $u_0>>1$ along with the high compression
approximation $u>>u_0$:
\begin{equation}
\dd{y}{u}=\frac{\gamma u^2}{\gamma -1}\,\exp[{\frac{T_0}{u\,u_0}}] \mbox{.}
\end{equation}
This equation still does not have an analytical solution but we can
add a term to the factor of the exponential to get the
derivative of the simple function $G(u,u_0)=u^4 \exp[{T_0}/(u\,u_0)]$.
Hence we finally take
\begin{equation}
\label{simple}
\dd{y}{u}=(1-4\frac{u\,u_0}{T_0})
\frac{\gamma u^2}{\gamma -1}\, \exp[{\frac{T_0}{u\,u_0}}]
\end{equation}
which will be a good approximation provided that $T_0>>u_0^2$.  This
yields the simple form $G(u,u_0)-G(u_a,u_0)$ for the function
$g(u,u_0)$ with which equation (\ref{trajg}) can then be integrated
numerically and tabulated to get a fast access to the shock
trajectory.

To check that the above approximations did not alter too much the
underlying cooling function, we can apply the procedure \ref{cool} to
the simplified equation (\ref{simple}). This provides:
\begin{equation}
\Lambda_1(\rho,p)=-\frac{T_0 
\left[ 1 + p\,\left( \gamma\,\rho-\gamma-1 \right)  \right]
(\rho -1) \rho }{\gamma (p-1) [4-4 p+T_0 (\rho
   -1)]}
\exp[{-\frac{T_0 (\rho -1)}{p-1}}]
\end{equation}

\begin{figure}
\centerline{ \psfig{file=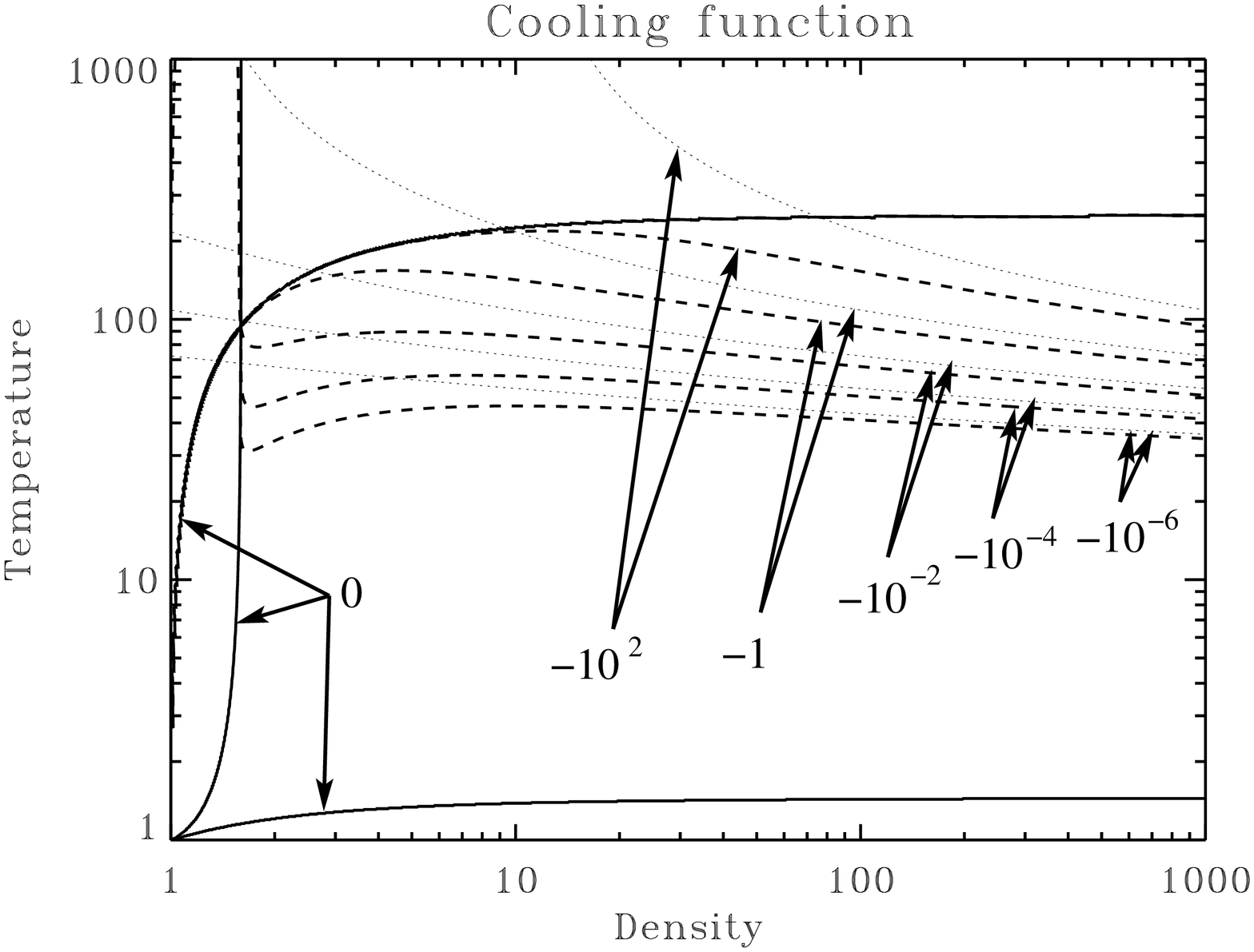,width=9cm,angle=-90} }
\caption{Contour plot of the cooling functions $\Lambda_0$ and
$\Lambda_1$ for $\beta=1$, $T_0=1000$ and $\gamma=5/3$
with respect to temperature $T=p/\rho$ and density $\rho$. Levels for
the contours are 0 (solid line, only for $\Lambda_1$) and -10$^2$, -1,
-10$^{-2}$, -10$^{-4}$ and -10$^{-6}$ from top to bottom for dotted
($\Lambda_0$) and dashed ($\Lambda_1$) lines. }
\label{lambda1}
\end{figure}

Figure \ref{lambda1} compares contour plots of both cooling functions
$\Lambda_0$ and $\Lambda_1$ (for $T_0=1000$ and $\gamma=5/3$). It can
be seen that despite the crude approximations we made, $\Lambda_0$ and
$\Lambda_1$ are still close to one another for a very large range of
parameters (for $1<T<<T_0$ and $\rho>>1$ both expressions asymptotically
converge). However, thermal equilibrium solutions (solid lines in
figure \ref{lambda1}) appear for $\Lambda_1$ when none existed for
$\Lambda_0$. Also, the range of applicable entrance velocities is
restricted to conditions such that the maximum temperature in the
shock is low compared to $T_0$ (because we made the $T_0>>u_0^2$
approximation).

This is nevertheless a good illustration of this method which can in theory
be applied to any cooling function. Indeed, one can in principle
use the fact that $u/u_0$ is bounded by a constant number strictly lower
than 1 to uniformly approach equation (\ref{statcool}) with
polynomials (or any other functional form simple to integrate). One
then recovers an analytical expression for $g(u,u_0)$ arbitrary
close to the exact solution. 

\section{Shock trajectory}
\label{shock}

\subsection{Exact solution}
Implicit ODEs like (\ref{trajg}) are in principle straightforward to
integrate numerically. It is however much harder to find an
analytically integrable form for these equations.  Such a solution
nevertheless exists for the simple but physically relevant example
(\ref{gfunc}).

 Let us use (\ref{gfunc}) into (\ref{trajg}) with $\beta=1$ to obtain

\begin{equation}
\label{rofrd}
r=\frac{2\gamma+(\gamma-1)v_0^2+(\gamma-3)\,\dot{r}\,v_0}
{(\gamma+1)\,(\dot{r}^2+\dot{r}\,v_0-1)}
\end{equation}

The solution of (\ref{rofrd}) for $\dot{r}$ yields only one positive
root $\dot{r}=h(r)$:

\begin{equation}
\label{rdofr}
h(r)=\frac{(\gamma-3)\,v_0 - (\gamma+1)v_0\,r+s(r)}{2(\gamma+1)\,r}
\end{equation}
where 
\begin{equation}
s(r)=\sqrt{a+b\,r+c\,r^2}
\end{equation}
with
\begin{equation}
a=(\gamma-3)^2\,v_0^2 \mbox{ ,}
\end{equation}
\begin{equation}
b=2(\gamma+1)[4\,\gamma+(\gamma+1)\,v_0^2] 
\end{equation}
and
\begin{equation}
c=(\gamma+1)^2(4+v_0^2)\mbox{.}
\end{equation}

 We are hence able to express the age $t$ of the shock
as a function of its position $r$ which provides the trajectory:

\begin{equation}
t(r)=\int_0^r \frac{{\rm d}x}{h(x)}
\end{equation}

We now write $1/h(x)$ as a sum of integrable terms:

\begin{equation}
\label{integrable}
\frac1{h(x)}=\frac12 v_0
+\frac{b+2c\,x}{4(\gamma+1)\,s(x)}
+\frac{e}{2s(x)}
-\frac{d}{2(z+x)}
+\frac{(\gamma+1)\,d^2}
{2(z+x)\,s(x)}
\end{equation}
 with
\begin{equation}
d=\frac{\gamma-1}{\gamma+1}v_0(3+v_0^2) \mbox{ ,}
\end{equation}
\begin{equation}
e=4\gamma+5(\gamma-1)v_0^2+(\gamma-1)\,v_0^4
\end{equation}
and
\begin{equation}
z=\frac{2\gamma}{\gamma+1}+\frac{\gamma-1}{\gamma+1}v_0^2\mbox{.}
\end{equation}

  We then integrate  expression (\ref{integrable}) to obtain the trajectory in
  its final analytical form 
$$ t(r)=
\frac{v_0}2 r
+\frac{s(r)-\sqrt{a}}{2(\gamma+1)}
+\frac{e}{2\sqrt{c}}
\log \left( \frac
{b+2 \sqrt{a\,c}}
{b+2c\,r+2\sqrt{c}\,s(r)}
\right)
+\frac{d}{2}\log \left( \frac{z}{z+r} \right)
$$\begin{equation}
+\frac{(\gamma+1)\,d^2}
{2s(-z)}
\log \left( \frac
{(z+r)(2a-b\,z+2\sqrt{a}\,s(-z))}
{z[(2a-b\,z)+(b-2c\,z)r+2s(r)s(-z)]}
\right)
\label{trajet}
\end{equation}

 If $r(t)$ is wanted, $t(r)$ can be numerically inverted by a Newton
 method of zero finding. This is done easily since the derivative
 $t'(r)=1/h(r)$ is provided analytically.

\subsection{High Mach number approximation}
\label{approx}
  We can recover a more simple but approximate trajectory if we 
make two assumptions :
\begin{itemize}
 \item in the limit of high Mach numbers ($v_0/c_0 \rightarrow
 \infty$), the adiabatic compression factor becomes a constant:
\begin{equation}
u_a(u_0)\simeq \frac{\gamma-1}{\gamma+1}u_0 \mbox{.}
\end{equation}
 \item at late times, the compression factor is nearly the isothermal
compression factor and $\dot{r}\simeq 1/u_0$. Hence for high Mach numbers
 $u_0=v_0+\dot{r}\simeq v_0$ and we use:
\begin{equation}
u_i(u_0)\simeq 1/v_0
\end{equation}

\end{itemize}

  With both these approximations, (\ref{gfunc}) with $\beta=1$ becomes
\begin{equation}
g(u,u_0)= v_0\frac{(1-\gamma)\,u_0+(\gamma+1)\,u}
{(\gamma+1)\,(u\,v_0-1)} \mbox{ ,}
\end{equation}
  the shock front velocity is:
\begin{equation}
 h(r)=\frac{(\gamma-1)\,v_0^2+(\gamma+1)\,r}
{v_0\,[2+(\gamma+1)\,r]}
\end{equation}
  and the resulting trajectory is
\begin{equation}
\label{lowM}
t(r)=v_0\,r+v_0\,(\frac{\gamma-1}{\gamma+1}v_0^2-\frac2{\gamma+1})
\log \left( \frac
{(\gamma-1)\,v_0^2}
{(\gamma-1)\,v_0^2+(\gamma+1)\,r}
\right) \mbox{.}
\end{equation}
  For early times (small $r$) the strong adiabatic shock trajectory
is recovered: 
\begin{equation}
\label{adiabatic}
t(r)\simeq \frac2{\gamma -1} \frac{r}{v_0} \mbox{.}
\end{equation}
  For very late times (very large $r$) the isothermal shock trajectory is
reached asymptotically:
\begin{equation}
\label{isothermal}
t(r)\simeq v_0\,r \mbox{.}
\end{equation}

\section{Numerical simulation}
\label{num}

\subsection{Numerical method}
  I compute here the time evolution of a radiative shock with cooling
  function (\ref{lnum}) thanks to a 1D hydrodynamical code. This code
  makes use of a moving grid algorithm \citep{DD87} which greatly
  helps to resolve the adiabatic front while keeping the total number
  of zones fixed to 100. The mesh driving function \citep[see][]{DD87}
  is designed to resolve the temperature gradients.

  The advection scheme is upwind, Donnor-Cell. The time integration is
  implicit fully non-linear with an implicitation parameter of 0.55 as
  a compromise between stability and accuracy. The time-step control
  keeps the sum of the absolute values of the variations of all
  variables lower than 0.5. In practice, the maximum variation of
  individual variables at each time-step is lower than 1\%.

I use a viscous pressure of the form:
\begin{equation}
p_v=\frac43 \rho c_s \sqrt{(\Delta x/10)^2+l^2} \max(-\dpart{u}{x},0)
\end{equation}
where $c_s=\sqrt{\gamma p/\rho}$ is the local sound speed, $\Delta x$ is the
local grid spacing and $l=10^{-3}$ is a prescribed dissipation length.

  To avoid numerical difficulties due to the form of $\Lambda(\rho,p)$
  when $p$ or $\rho$ are close to 1, we set $\Lambda$ to zero when
  $p-1$ or $\rho-1$ are lower than 10$^{-3}$. The time normalisation
  is such that $\beta=1$.

  The entrance parameters are set to $(\rho,p,v_0)=(1,1,10)$ with
  $\gamma=5/3$ and the evolution is computed until a stationary state
  is nearly reached.

\subsection{Trajectory}
\label{numtraj}
\begin{figure}
\centerline{
\epsfig{file=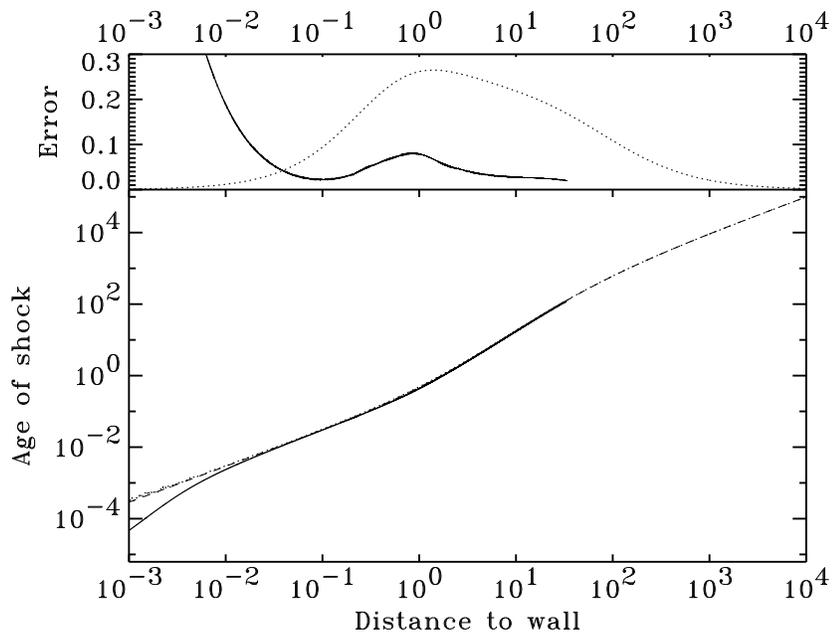,width=9cm,angle=90}
} \caption{Trajectory of the shock age vs position in the simulation
(solid) compared to the analytic expression for a quasi-steady shock
(dashed) and the high Mach number approximation (dotted). Also shown
(solid curve in upper pannel) is the relative error between the solid
and dashed curves as well as the estimate $\ddot{r}t^2/r$ for this
error given in section \ref{discussion} (dotted curve in upper
pannel).}
\label{traj}
\end{figure}

  The position of the shock at each time-step is computed as the
  position of the maximum of the ratio $p_v/p$ along the simulation
  box.  I compare this trajectory to the analytical expression
  (\ref{trajet}) on figure \ref{traj}.  At a given position, the
  relative difference on the ages of the shock is maximum at the very
  beginning, when the shock front is being formed and the position is
  still no more than a few dissipative lengths.  For times greater
  than $5\times10^{-3}$, the relative error is less than 8\%, with a
  secondary maximum at $r\simeq 1$. An estimate for this error is
  given in section \ref{discussion}.  Note that both the isothermal
  (\ref{isothermal}) and the adiabatic (\ref{adiabatic})
  approximations are wrong by an order of magnitude at this point. The
  adiabatic approximation is accurate to a 20\% level only for ages
  lower than 0.1. Afterward the effects of cooling slow down the shock
  and the adiabatic solution overestimate the position by large. The
  isothermal approximation is valid up to a 20\% level only for times
  greater than 3000. This is because it does not take into account the
  period of early times when the shock is moving swiftly and since the
  isothermal shock moves at a slow pace it takes time to recover the
  delay. In other words, the cooling history of the shock does make a
  difference to its position. Since the ratio between the adiabatic
  speed and the isothermal speed scales like $v_0^2$ this situation
  will be even worse for stronger shocks. By contrast, the error
  estimate given in section \ref{discussion} suggests that the
  quasi-steady approximation is equally good for stronger shocks.

  The approximate trajectory for high Mach number quasi-steady shocks
  (\ref{lowM}) is also shown. It is already a very good approximation
  even for the relatively low Mach number ($v_0/c_0=7.75$) I use. Indeed in
  section \ref{approx} I neglected only terms of order greater than 2 in the
  inverse of the Mach number.

\subsection{Snapshots}
  I output the results of the simulations at a few selected
  time-steps.  For each of these snapshots, I determine the position
  $r$ of the shock with $p_v/p$ as in the previous subsection
  \ref{numtraj}. I then compute the velocity of the quasi-steady
  shock front $\dot{r}=h(r)$ thanks to
  (\ref{rdofr}). $u_0=v_0+\dot{r}$ gives the entrance velocity in the
  frame of the steady shock. I now recover the relation $u=f(y,u_0)$
  thanks to equations (\ref{steadyg}) and (\ref{gfunc}):
\begin{equation}
u=\frac{-2\gamma+(1-\gamma)\,u_0^2+(\gamma+1)y}{(\gamma+1)\,u_0(y-1)} \mbox{.}
\end{equation}
  The temperature ($T=p/\rho$) profile can finally be retrieved from this 
velocity profile thanks to the relations (\ref{smass}) and (\ref{smom}).

\begin{figure}
\centerline{
\psfig{file=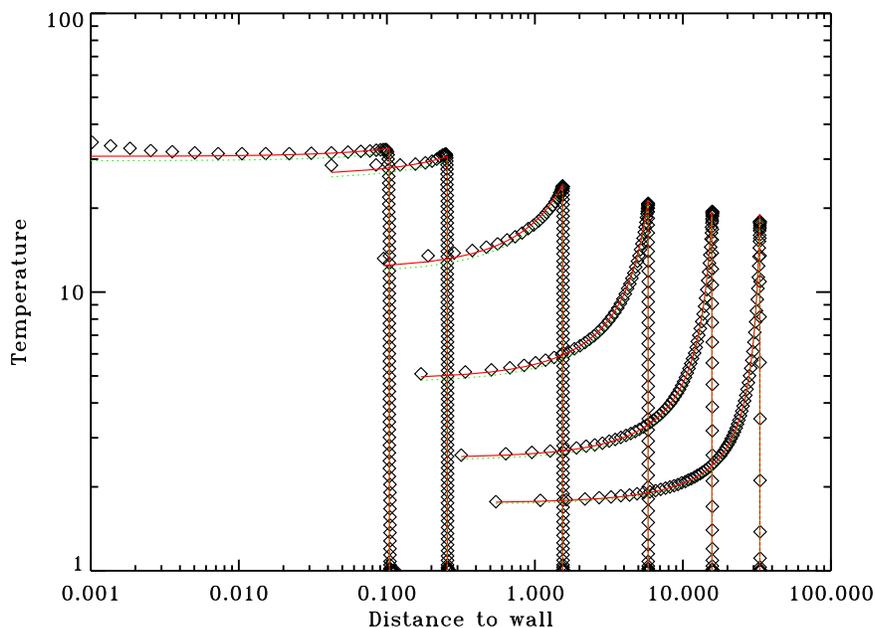,width=9cm,angle=+90}
} 
 \caption{Temperature ($T=p/\rho$) profiles in the hydrodynamical
simulation (diamonds) compared to the analytical solution for a
quasi-steady shock (solid red line) and its high Mach number approximation
(dotted green line).  }
\label{snap}
\end{figure}

I compare the quasi-steady state solution to the results of the
numerical simulation on figure \ref{snap}. The gas is flowing from the
right onto the wall on the left. At early times, when the shock front
is still close to the wall, the temperature at the wall is higher in
the simulation than in the quasi-steady shock. This is mainly due to
the wall heating effect, which decreases at later times when the
cooling function has a stronger influence on the temperature. The
decrease of the maximum temperature in the shock is due to the
decrease of the relative entrance velocity of the gas in the adiabatic
shock front (see figure \ref{snap}). Note the high resolution provided
by the moving mesh at the adiabatic shock front. For later times at
the end of the relaxation layer, the temperature decreases toward its
final value of $T=1$ at equilibrium.

The dotted curves in figure \ref{snap} are the high Mach number solutions
(see section \ref{approx}). They already are very close to the exact solutions
as the approximation is of order 2. 

\section{Discussion}

\subsection{Time-dependent shocks}
\label{discussion}

  The differences between the quasi-steady state and the numerical
  solution described in section \ref{num} both come from the numerical
  errors in the scheme and from the fact that quasi-steady shocks are
  only an approximation to time-dependent shocks. I give here an estimate
 on the difference between quasi-steady shocks and shocks.
  
  I now write the inviscid equations of time-dependent
  hydrodynamics that a time-dependent shock would set to zero:
\begin{equation}
E_{\rho}=\dpart{\rho}{t}+\dpart{(\rho\, v)}{x} \mbox{ ,}
\end{equation}
\begin{equation}
E_v=\dpart{v}{t}+v\,\dpart{v}{x}+\frac1{\rho}\,\dpart{p}{x}
\end{equation}
and 
\begin{equation}
E_p=\frac1{(\gamma-1)}\,\dpart{p}{t}
+\frac1{(\gamma-1)}\,\dpart{(p\, v)}{x}+p\,\dpart{v}{x}-\Lambda(\rho,p) 
\end{equation}
 where $\rho(x,t)$, $v(x,t)$ and $p(x,t)$ are the density, velocity
 and pressure fields in the frame of the wall.
  
  If we now use the quasi-steady solutions to express these equations,
we find
\begin{equation}
\label{fail1}
E_{\rho}=\dpart{\rho_s}{u_0}\,\ddot{r} \mbox{ ,}
\end{equation} 
\begin{equation}
E_v=(\dpart{u_s}{u_0}-1)\,\ddot{r}
\end{equation}
 and 
\begin{equation}
\label{fail3}
E_p=\frac1{(\gamma-1)}\dpart{p_s}{u_0}\,\ddot{r}
\end{equation}
 where $\rho_s(y,u_0)$, $u_s(y,u_0)$ and $p_s(y,u_0)$ are the solutions
of the steady state equations in the frame of the shock.
 
Hence quasi-steady shocks are in general only approximations to
time-dependent shocks. The quasi-steady state approximation amounts to
neglecting the acceleration $\ddot{r}$ of the shock front.  Note that
a maximum departure of the trajectory from the numerical simulation
occurs around $r \simeq 1$ when the shock switches from adiabatic
velocities to isothermal velocities (see figure \ref{traj}), ie: when
accelerations are likely to be the highest. A rough estimate for
the relative error on the position is hence $\ddot{r}t^2/r$ which
overestimates the error by more than about a factor 3 (see figure
\ref{traj}). Interestingly, this estimate does not depend on the shock
velocity for strong shocks. 

 It is also interesting to note from equations
 (\ref{fail1})-(\ref{fail3}) that a high dependence of the steady
 state on the entrance velocity $u_0$ will cause departures of
 time-dependent shocks from the quasi-steady state.  In this context,
 refer to \cite{PII} who found that the quasi-steady state was
 violated for marginally dissociative shocks. For this type of shocks,
 the entrance velocity is indeed close to the critical velocity at
 which the major cooling agent is dissociated and a small variation
of the entrance velocity can strongly affect the post-shock.

\subsection{Entrance parameters}
 In section 3, I used a normalisation based on the entrance values for
 the density and pressure: this must not hide the fact that the
 cooling function (\ref{lnum}) implicitly depends on these parameters.
 Hence for a given cooling function, I computed analytical solutions
 only for a fixed set of entrance density and pressure. At this point,
 there is no reason why other sets of parameters should provide
 integrable solutions.

\section{Summary and future prospects}
  I described a general way of obtaining time-dependent analytical
  solutions to quasi-steady shocks with cooling. I applied this method
  to a physically sensible example and compared the resulting
  quasi-steady shock to a time-dependent hydrodynamic simulation.  I
  also provided a more simple high Mach number approximation to the
  exact solution. I showed that even though quasi-steady shocks are
  not strictly speaking time-dependent shocks, they are a good
  approximation to time-dependent shocks. In particular, more simple
  approximations such as the adiabatic approximation or the
  steady-state approximation badly fail to reproduce the behaviour of
  the shock for a large range of times. This is because the cooling
  history of the shock is essential in determining the position of the
  shock.

  I wish to emphasise also that the method described in section
  \ref{cool} allows to recover the underlying cooling function for any
  set of steady state solutions. The associated quasi-steady shocks
  can then be easily computed by solving the ODE~(\ref{trajg}). As
  demonstrated in subsection \ref{apcool} one can fit any given
  cooling function with analytical forms for the functions
  $g(u,u_0)$. This hence provides a potentially powerful method to
  quickly compute the evolution of any quasi-steady shock with
  cooling.

  Analytical solutions of quasi-steady shocks can be used as a basis
  to study properties of time-dependent shocks. Linear analysis around
  the quasi-steady solution may provide insight for the time-dependent
  behaviour of shocks. They can also help to address under what
  conditions shocks tend toward the quasi-steady state.

  Furthermore, this work represents a first step towards exact
  solutions or better approximations to time-dependent problems of
  non-adiabatic hydrodynamics. In the future this might lead to new
  algorithms for dissipative hydrodynamical simulations.  Finally,
  note that similar procedures can be applied to shocks with chemistry
  and magnetic fields \citep[using the results of][]{PII}.

\section*{Acknowledgements}
  Many thanks to Dr Neil Trentham for introducing to me Mathematica,
  which made this work a lot easier.



\end{document}